\documentclass{aa}  

\usepackage{graphicx}
\usepackage{txfonts}
\usepackage[]{hyperref}
\usepackage{subcaption}
\usepackage{graphicx}

\begin{document}

   \title{The stellar population of a $z\sim3.25$ Ly$\alpha$ emitting group associated with a damped Ly$\alpha$ absorber}

    \titlerunning{Stellar population near a $z\sim3.25$ DLA.}

   \author{
            Giulia Pruto\inst{1,2}\fnmsep\thanks{g.pruto@campus.unimib.it},
            Michele Fumagalli\inst{1,3}\fnmsep\thanks{michele.fumagalli@unimib.it},
            Marc Rafelski\inst{4,5},
            Mitchell Revalski\inst{4},
            Matteo Fossati\inst{1},
            Ruari Mackenzie\inst{6}
            \and
            Tom Theuns\inst{7}
            }

\authorrunning{Giulia Pruto et al.}

   \institute{Dipartimento di Fisica G. Occhialini, Universit\`a degli Studi di Milano Bicocca, Piazza della Scienza 3, 20126 Milano, Italy
        \and
            Institute for Astronomy, University of Edinburgh, Blackford Hill, Edinburgh, EH9 3HJ, UK 
         \and
             INAF – Osservatorio Astronomico di Trieste, via G. B. Tiepolo 11, I-34143 Trieste, Italy
        \and
             Space Telescope Science Institute, 3700 San Martin Drive, Baltimore, MD 21218, USA
        \and 
            Department of Physics and Astronomy, Johns Hopkins University, Baltimore, MD 21218, USA
        \and
            Department of Physics, ETH Zürich, Wolfgang-Pauli-Strasse 27, Zürich, 8093, Switzerland,
        \and
            Institute for Computational Cosmology, Department of Physics, Durham University, South Road, Durham DH1 3LE, UK\\
            }

   \date{\today}

\abstract{We present near-infrared observations, acquired with the Wide Field Camera 3 (WFC3) on board of the {\it Hubble Space Telescope} (HST),  of a Ly$\alpha$ double-clumped emitting nebula at $z \approx 3.25$ associated with a damped Ly$\alpha$ absorber (DLA). 
With the WFC3/F160W data we observe the stellar continuum around $3600$~\AA\ in the rest frame for a galaxy embedded in the West clump of the nebula, $G_{\rm W}$, for which we estimate a star formation rate SFR$_{G_{\rm W}} = 5.0 \pm 0.4$~M$_\odot$~yr$^{-1}$ and maximum stellar mass M$_{G_{\rm W}} < 9.9 \pm 0.7 \times 10^9$~M$_\odot$. With the enhanced spatial resolution of HST, we discover the presence of an additional faint source, $G_{\rm E}$, in the center of the East clump, with a star formation rate of SFR$_{G_{\rm E}} = 0.70 \pm 0.20$~M$_\odot$~yr$^{-1}$ and maximum stellar mass M$_{G_{\rm E}} < 1.4 \pm 0.4 \times 10^9$~M$_\odot$. We show that the Ly$\alpha$ emission in the two clumps can be explained by recombination following {\it in-situ} photoionization by the two galaxies, assuming escape fractions of ionizing photons of $\lesssim 0.24$ for $G_{\rm W}$ and $\lesssim 0.34$ for $G_{\rm E}$. The fact that $G_{\rm W}$ is offset by $\approx 8$~kpc\ from the West clump does not fully rule out the presence of additional fainter star-forming sources that would further contribute to the photon budget inside this $\approx 10^{12}$~M$_\odot$ galaxy group that extends over a region encompassing over $30 \times 50$~kpc. 
}

\keywords{Galaxies: groups: general -- Galaxies: halos -- Galaxies: high-redshift -- quasars: absorption lines}

\maketitle
%

\section{Introduction}

Spectroscopy with large-format integral field unit (IFU) on 8m class telescopes has rapidly transformed our view of the circumgalactic medium (CGM), the heterogeneous diffuse gas around galaxies that plays a key role in the regulation of the baryon cycle. Once the realm of absorption spectroscopy alone, the study of the CGM has rapidly progressed thanks to the Multi Unit Spectroscopic Explorer (MUSE) \citep{Bacon2010} at the Very Large Telescope (VLT) and other IFUs such as the Keck Cosmic Web Imager (KCWI) \citep{Morrissey2018}. With these instruments, halo gas is mapped routinely in Ly$\alpha$ emission up to scales of $\approx 100~$kpc near star-forming galaxies \citep[][]{Wisotzki2016,Leclercq2017} and quasars \citep{Borisova2016, ArrigoniB2019, Fossati2021}, metals in emission have been imaged in the inner CGM of both active galactic nuclei and star-forming systems \citep[][]{Guo2020, Fossati2021, Dutta2023, Guo2023}, cosmic web filaments connecting multiple galaxies have been unveiled in low surface-brightness Ly$\alpha$ maps \citep[][]{Umehata2019,Bacon2021}.
 
Unlike traditional techniques such as multiobject spectroscopy, employed, for example, in the Keck Baryonic Structure Survey \citep[KBSS,][]{Rudie2012KBSS, Steidel2014KBSS} and in the VLT Lyman break galaxy Redshift Survey \citep[VLRS, e.g.,][]{Shanks2011VLRS, Bielby2011VLRS, Crighton2011VLRS}, IFUs do not require any pre-selection of galaxies. 
This advantage has made IFUs a common tool for studying galaxies associated with absorption line systems, uncovering many sources at smaller impact parameters that went undetected when observed with other instruments.
Furthermore, thanks to dedicated surveys \citep[e.g.,][]{Schroetter_Bouche2016, Peroux2019, Lofthouse2020, Muzahid2020, Oyarzun2024} targeting continuum faint Ly$\alpha$ or [OII] emitters, samples of absorbers and galaxy associations have increased in size from a few tens to several thousands. At $z\approx 0.5-1.5$, the MAGG survey \citep{Dutta2020} has expanded the results of multi-object spectrographs \citep[][]{Chen2018,Weiner2009}
to higher redshifts ($z\approx 1.5$), revealing that stellar mass is the dominant factor influencing the \ion{Mg}{II} absorption around galaxies.
By focusing on the inner CGM traced by very strong \ion{Mg}{II} absorption systems, the MEGAFLOW survey \citep{Schroetter_Bouche2016}
has provided an expanded view of the effects of inflows and outflows on the column density and kinematic distributions of the absorbing gas around the host galaxy as a function of the galaxy orientation. At higher redshift, $z\gtrsim 3$, the MUSEQuBES \citep{Muzahid2020} and MAGG surveys 
\citep[][]{Lofthouse2020, Lofthouse2023, Galbiati2023} have extended our view of the properties of hydrogen and metals (traced by \ion{C}{IV} and \ion{Si}{IV}) around continuum-faint Ly$\alpha$ emitters (LAEs), reaching $\approx 1~$dex lower mass compared to previous studies that used brighter Lyman break galaxies (LBGs). These studies have revealed the presence of gas filaments hosting strong hydrogen and metal absorbers stretching across galaxies, as well as diffuse pockets of lower column density and enriched gas.

The analysis of the galactic environment around metal absorbers has been pushed to even higher redshift ($z\gtrsim4$) thanks to the combined power of the NIRCam slitless grism spectrograph on board of the James Webb Space Telescope (JWST) and the Atacama Large Millimeter Array (ALMA). 
These studies revealed the presence of galaxies at impact parameter $< 300$ kpc from low-ionization metal absorbers, suggesting the presence of an efficient IGM enrichment mechanism during the later stages of reionization \citep[e.g.,][]{Wu2023, Bordoloi2024}.

Thanks to integral field spectroscopy, it has clearly emerged that absorption line systems are often associated with multiple galaxies, including cases of rich galaxy groups with up to $\approx 10$ members. 
These rich groups present more extended distributions of both hydrogen and metal-enriched gas 
\citep[][]{Bordoloi2011,Fossati2021, Dutta2020, Galbiati2023, Lofthouse2023, Muzahid2021}, with covering factors $\approx 2-5$ times higher than those of isolated galaxies. While the mechanisms responsible for this elevated gas distribution remain at present unconstrained, the study of individual cases where tomography in absorption is possible or enriched gas in emission can be probed 
\citep[][]{Fossati2019,Chen2019,Leclerq2022} point to gravitational interactions and outflows as possible mechanisms that can increase the contribution arising from the superposition of halos or a more diffuse intragroup medium.  

Among the first examples of absorption line systems associated with group environments observed by MUSE, \citet{Fumagalli2017} reported the detection of a damped Ly$\alpha$ absorber (DLA) with column density 
of $\mathrm{log} \ N_{\mathrm{HI}} = 20.85 \pm 0.10$ cm$^{-2}$ at redshift $z_{\mathrm{dla}} =  3.2552 \pm 0.0001$. This DLA is associated with a UV-continuum detected galaxy at a projected distance of $19.1 \pm 0.05$ kpc, embedded in a Ly$\alpha$ extended nebula composed of two bright clumps, separated by a projected distance of $16.5 \pm 0.5$ kpc.
The line of sight velocity of the two Ly$\alpha$ emitting clumps is aligned in velocity with the main absorption components of metal lines associated with the DLA, which suggests a link between the absorption and emission sub-structures. 
This evidence is consistent with multiple galaxies forming inside an extended gas-rich and metal-rich structure.
As the two clumps were detected only via Ly$\alpha$, uncertainty remained about the actual nature of this emitting region and the possible powering mechanisms. 

To address these questions, we have collected near-infrared (NIR) imaging (PID 15283; PI Mackenzie) using the F160W filter with Wide Field Camera 3 (WFC3)  
on board the {\it Hubble Space Telescope} (HST) to search for rest-frame optical emission and better constrain the properties of this system through stellar population synthesis analysis.

This paper discusses the HST NIR follow-up observations and expands on the conclusions presented in \citet{Fumagalli2017}. The structure of the paper is as follows. In Sect.~\ref{obs_datareduction} we present the new observations and data reduction, followed by the analysis in Sect.~\ref{sec:analysis} and a discussion on the nature of the Ly$\alpha$ emission origin and the associated galaxy environment in Sect.~\ref{sec:discussion}. The summary and conclusions are presented in Sect.~\ref{sec:conclusions}. Throughout, unless otherwise noted, we quote magnitudes in the AB system, distances in proper units, and adopt the Planck 2015 cosmology \citep[$\Omega_{\mathrm{m}} = 0.307$, $H_0 = 67.7$~km~s$^{-1}$~Mpc$^{-1}$;][]{Planck2016}. 

\section{MUSE and HST observations}
\label{obs_datareduction}

\subsection{Spectroscopy from MUSE}

The quasar J0255+0048 was first observed thanks to an imaging survey aiming to probe in situ star formation associated with DLAs \citep{Omeara2006, Fumagalli2010}.
From this survey, J0255+0048 and other five quasars were selected as systems hosting DLAs at $z>3$, the redshift for which Ly$\alpha$ enters the wavelength range covered by MUSE, allowing the gathering of additional data.
MUSE observations of these six quasar fields were conducted at the VLT as part of the ESO programs 095.A-0051 and 096.A-0022 (PI Fumagalli).
Observations were carried out on the nights of 17-20 September 2015 in a series of 1500~s exposures, with a total of 2.5~hours under good seeing conditions (requested to be $\leq 0\farcs8$) and clear sky.

The detailed data reduction process is presented in \citet{Fumagalli2017}; only the key steps are highlighted here. Following the standard ESO MUSE pipeline \citep[v1.6.2;][]{Weil2014}, basic corrections such as bias subtraction and flat fielding were applied in addition to wavelength and photometry calibrations. The frames were then processed to improve the quality of sky subtraction and flat-fielding and to remove the residuals left from the ESO pipeline reduction using the \textsc{CubExtractor} code \citep[\textsc{CubEx}, ][]{Cantalupo2019}. Corrections for extinction were implemented. From comparisons with photometric data from the Sloan Digital Sky Survey \citep{Eisenstein2011} a factor of 1.12 was applied to the flux calibration to consider low levels of atmospheric extinction. Following \citet{Schlafly2011} the presence of Milky Way dust in the direction of observation was evaluated.

The final IFU cube has a field of view of $1\times 1$~arcmin$^2$ composed by 0$\farcs$2\ pixels, covering the wavelength range 4750-9350~\AA\ in bins of 1.25~\AA. At $\lambda \approx 5170 \ \AA$, corresponding to the Ly$\alpha$ wavelength at the DLA redshift, the effective image quality is $\approx 0\farcs6$ and the noise level $\approx 6 \times 10^{-19}$~erg~s$^{-1}$~cm$^{-2}$~$\AA^{-1}$~arcsec$^{-2}$ (root mean square, RMS).

\begin{figure*}
    \centering
    \includegraphics[width=15cm]{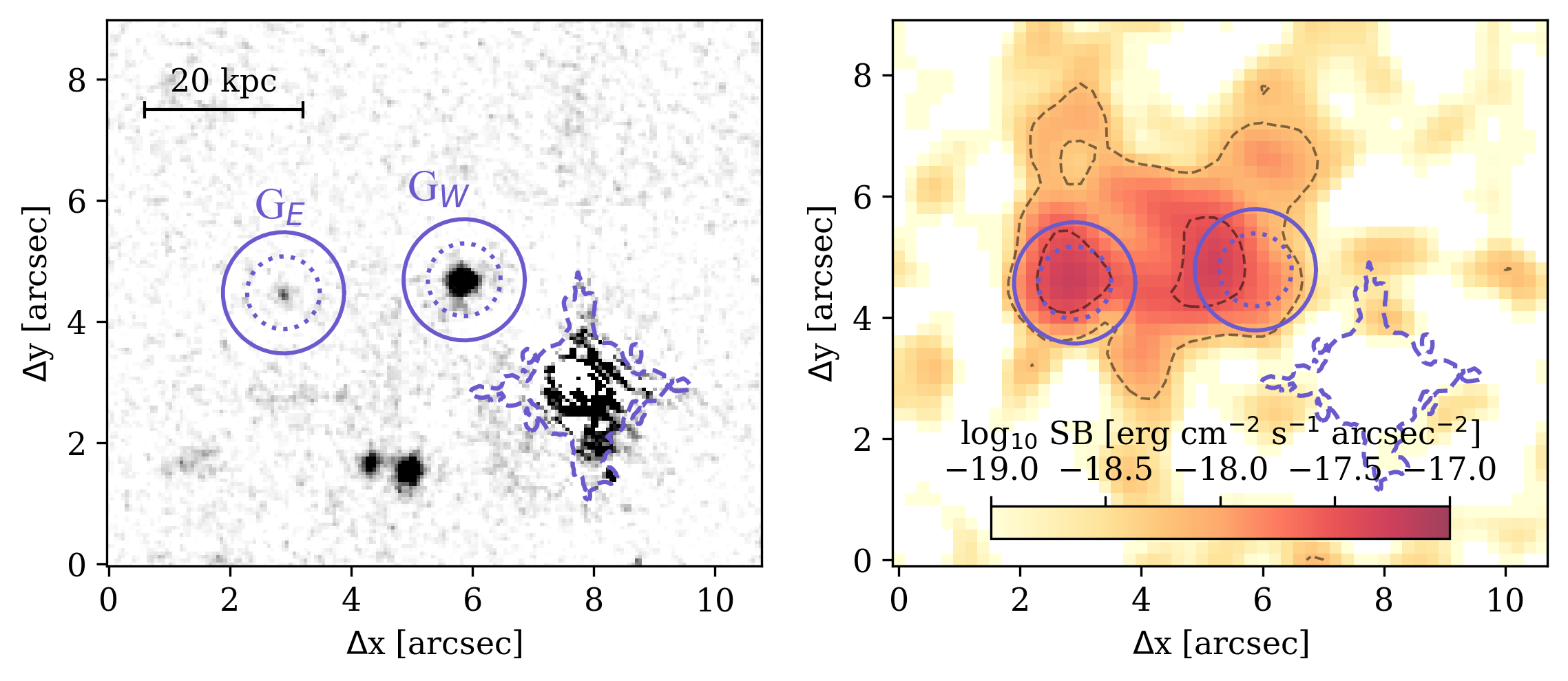}
    \caption{\textit{Left}: HST WFC3/F160W observation with the subtracted quasar PSF. The two sources $G_{\rm E}$ and $G_{\rm W}$ are indicated by the purple circular apertures of $0\farcs6$ radius (dotted line) and $1\arcsec$ (solid line) centered on the galaxies. Next to $G_{\rm W}$, the dashed purple line marks the original PSF of the quasar.
    \textit{Right}: Ly$\alpha$ surface brightness map obtained from the MUSE cube and smoothed with a Gaussian 2D kernel with $\sigma_x = \sigma_y = 0\farcs3$. The dashed gray lines are delimiting the $1\sigma$ ($6\times 10^{-19}$~erg~s$^{-1}$~cm$^{-2}$~arcsec$^{-2}$) and $5\sigma$ surface brightness levels. The purple circles of radius $0\farcs6$ and $1\arcsec$ surrounding the continuum detected sources are reproduced to better show the positions of the two galaxies relative to the Ly$\alpha$ clumps ($C_{\rm W}$ and $C_{\rm E}$), and the position of the quasar is again marked by the purple dashed line.}
    \label{fig:position_Lya}
\end{figure*}

\subsection{Near infrared imaging from HST}

The field of quasar J0255+0048 was imaged over 4 orbits with HST WFC3/IR F160W (PID 15283; PI Mackenzie).
Observations were obtained on 28 August 2018 and 15 September 2019. The bright quasar has $H(AB)=18.33$~mag, several magnitudes brighter than the targeted nebula counterparts at a separation of $2-6\arcsec$. Given this magnitude contrast, ensuring that the quasar did not contaminate the structure either through diffraction spikes or by dithering the structure onto pixels affected by persistence from the bright quasar was critical. To minimize the impact of diffraction spikes, we selected only ORIENT angles that place the diffraction spikes away from the structure identified in Ly$\alpha$ emission. 

There is a bright star with $H(AB)\approx 9.0$~mag $\approx 45\arcsec$ away from the quasar and the Ly$\alpha$ nebula. The star is sufficiently bright that it would quickly saturate the detector and its extended point spread function (PSF) would contaminate the object of interest. We, therefore, offset the exposures from the default aperture so that the star falls outside the field of view to the greatest extent possible. This resulted in our target being placed in one corner of the observed field, with diffraction spikes from the star that do not overlap with our target source.

To control persistence and dither to improve spatial sampling, we adopted the LINE dither pattern with 1.5\arcsec\ point spacing to move along the line separating the quasar and the Ly$\alpha$ nebula. In this way, the structure never fell on pixels that the quasar had fallen on in previous exposures. The larger point spacing ($\approx3 \times$) allows the removal of IR blobs and other artifacts for cleaner images.  To maximize the signal-to-noise ratio ($S/N$) for any faint counterparts of the Ly$\alpha$ nebula, we used SPARS50 with NSAMP=14, resulting in 4 exposures per orbit over 3 orbits. 

The observations in 2019 were repeated because observations from 2018 failed due to a guide star re-acquisition failure, resulting in the loss of an exposure. The orientation constraints required to avoid the diffraction spikes delayed the repeat for a year. This provided an opportunity to further avoid the diffraction spikes of the bright nearby star for a more fully cleaned image. Thus, we slightly offset the target location of this repeated visit. We obtained an additional 4 exposures over one orbit with the same dither pattern and sampling as the original data, although at a different orientation ($U3=272.2^{\circ}$ compared to $U3=274.0^{\circ}$ for the original visit). In total, we obtained $9,794$~s of successful exposure time over 15 dithered exposures.

The data were downloaded from MAST in 2022 to obtain data calibrated with the IR filter-dependent delta sky flats by date of appearance of IR blobs (WFC3 ISR 2021–01; \citealt{Olszewski2021}). Custom masks were created to remove the diffraction spikes from the nearby star and satellite trails. The individual exposures were first aligned to each other with \textsc{TweakReg}. Then the resultant combined image mosaic generated with \textsc{AstroDrizzle} was aligned using
\href{https://www.cosmos.esa.int/web/gaia/early-data-release-3}{Gaia~EDR3} \citep{GaiaCollaboration2021} with uncertainties $\lesssim 0\farcs04$.  
The following drizzle parameters were used: \textsc{combine\_type} was set to ``imedian'', \textsc{skymethod} to ``globalmin+match'', \textsc{final\_wht\_type} to ``IVM'', \textsc{final\_scale} to 0.06,  \textsc{final\_pixfrac} to 0.8, and \textsc{final\_rot} to 0. 

RMS error images were created from the resultant weight map (WHT; inverse variance map), where $RMS=1/\sqrt{WHT}$. We corrected the RMS maps for correlated pixel noise as described in \S3.3.2 of the \textsc{DrizzlePac} Handbook (v2.0, \citealt{Hoffmann2021}), which provides a noise scaling factor ($R$) based on the drizzled and native pixel scales, and the \textsc{final\_pixfrac}. For our drizzle parameters, the noise scaling factor is $R$~=1.71, yielding a correlated noise factor of 2.124.

\section{Analysis of HST imaging data}
\label{sec:analysis}

With HST NIR follow-up observations, two sources, labeled $G_{\rm E}$ and $G_{\rm W}$, were detected in emission in the region of the double-clumped Ly$\alpha$ nebula.
Here, $G_{\rm W}$ is the galaxy previously named $G$ in \citet{Fumagalli2017}.
The two sources are separated by a projected distance of $2\farcs99$, $\approx 23$ kpc at the DLA redshift $z_{\rm dla}$, with a projected distance between $G_{\rm E}$ and the quasar of $\approx 40$ kpc and between $G_{\rm W}$ and the quasar of $\approx 19.5$ kpc.

From the comparison between the Ly$\alpha$ emission and the HST NIR observations in Fig.~\ref{fig:position_Lya}, we find that a third ($\approx 33 \%$ in the case of $G_{\rm W}$ and $\approx 27\%$ for $G_{\rm E}$) of the total Ly$\alpha$ emission of the nebula is detected inside the $1\arcsec$ circular apertures around the two sources (solid purple circles), suggesting a connection between the star formation activity within the detected galaxies and the Ly$\alpha$ emission.
Moreover, when comparing the centers of the two Ly$\alpha$ emitting clumps, C$_{\rm W}$ and C$_{\rm E}$, with the centers of the two galaxies detected in the continuum, we find an offset of $\approx 7.7$~kpc between $G_{\rm W}$ and $C_{\rm W}$ and $\approx 1.6$~kpc between $G_{\rm E}$ and $C_{\rm E}$.
$G_{\rm W}$ has a spectrum and colors consistent with those of a $z \approx 3.2$ Lyman break galaxy (LBG) \citep{Fumagalli2017}. We do not have previous information about the second galaxy, $G_{\rm E}$, because its fainter continuum emission is not detected by MUSE. However, the fact that $G_{\rm E}$ is well centered in correspondence of the Ly$\alpha$ emitting clump $C_{\rm E}$ is indicative of a true association rather than a spurious projection effect. For this reason, the redshift of both galaxies is assumed to be that of the DLA, $z_{\rm dla} = 3.2552 \pm 0.0001$.
At the estimated redshift of these galaxies, the [\ion{O}{III}] and H$\alpha$ lines could be observed at wavelengths much higher than the ones covered by the F160W filter. 
Other lines that can fall in the observed wavelength range generally have lower equivalent widths, and for this reason no contributions from emission lines is included.

\subsection{Photometry}

Flux from the bright quasar J0255+0048 needs to be subtracted to accurately compute the flux emitted by the two sources. The quasar contribution within the circular aperture of $1\arcsec$ around $G_{\rm E}$ is consistent with zero within the background uncertainty and, therefore, negligible. 
For $G_{\rm W}$, a subtraction of the quasar PSF  with a bespoke code developed for HST/WFC3 PSF modeling \citep{Revalski2022, Revalski2023} suppressed the quasar contamination in the circular area of $1\arcsec$ radius around $G_{\rm W}$ to levels comparable with zero given the background uncertainty. 
All the background uncertainties related to a given aperture in the science image are evaluated as the square root of the quadrature sum of the values displayed by the pixels of the RMS image inside the same aperture.

\begin{figure*}
    \centering
    \begin{subfigure}{0.45\textwidth}
        \includegraphics[width=\textwidth]{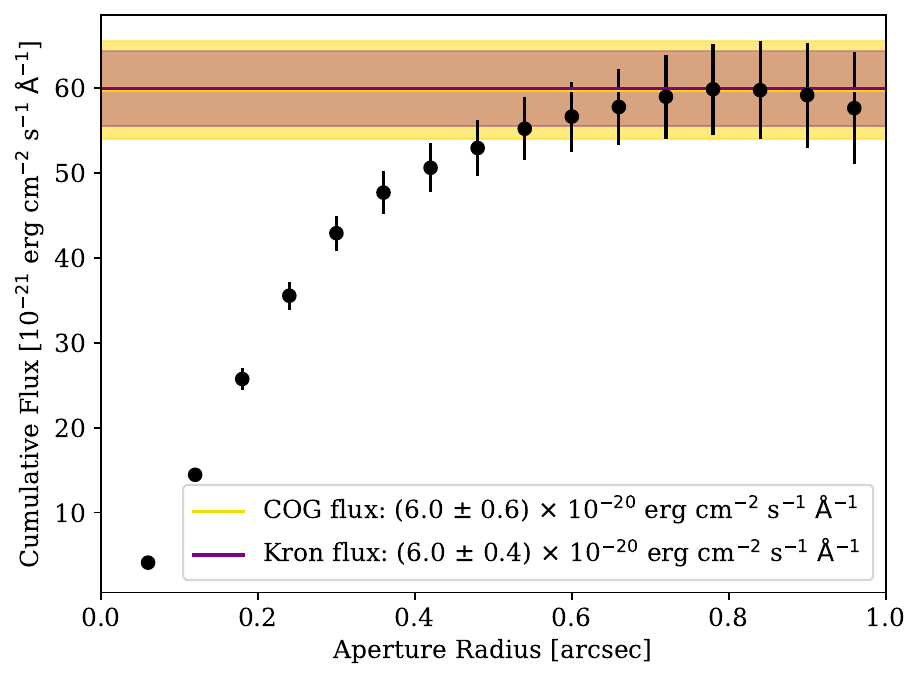}
        \caption{G$_{\rm W}$}
        \label{fig:cog_GW}
    \end{subfigure}
    \begin{subfigure}{0.45\textwidth} 
        \includegraphics[width=\textwidth]{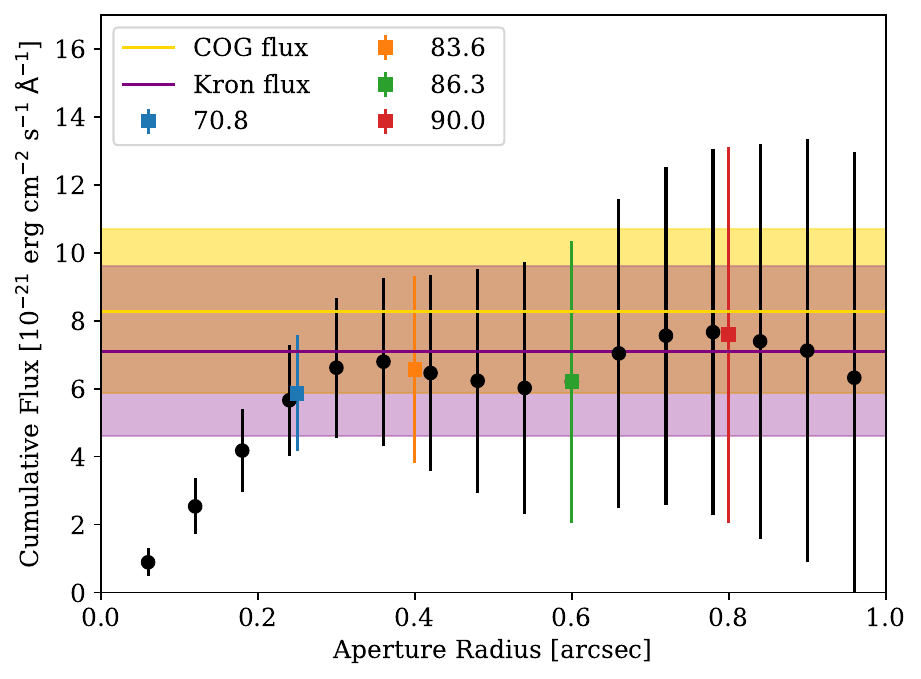}\caption{G$_{\rm E}$} 
        \label{fig:cog_GE}
    \end{subfigure}
    \caption{(a) Growth curve for galaxy $G_{\rm W}$. The black points show the resulting flux with background and shot-noise uncertainties. 
    The Kron flux is represented in purple with the shaded 68\% confidence region. The growth curve flux is marked in yellow with the shaded 68\% confidence region.
    Excellent agreement is found comparing the two estimates. (b) Same as the left panel but for source $G_{\rm E}$. The colored squares represent the flux observed inside the circular apertures that encircle  70.8\%, 83.6\%, 86.3\%, and 90.0\% of the total energy emitted by a point-like source.}
    \label{fig:cog}
\end{figure*}

To measure the flux emitted by $G_{\rm W}$ we use the \textsc{Source Extractor} Python library \citep{Bertin1996,Barbary2018} and compute the signal contained inside the Kron elliptical aperture associated to the galaxy, expected to encircle 94\% of the total flux \citep{Kron1980, Bertin1996}. The measured value is then divided by a factor of 0.94 to derive the total flux.
Fig.~\ref{fig:cog_GW} shows in purple the measured Kron flux for $G_{\rm W}$, $f_{G_{\rm W}, \rm Kron} = 6.0 \pm 0.4 \times 10^{-20}$~erg~s$^{-1}$~cm$^{-2}$~$\AA^{-1}$.
The corresponding AB magnitude is $m(AB)_{G_{\rm W}} = 24.71 \pm 0.08$~mag.
Uncertainties on flux measured inside a given aperture are evaluated as the square root of the quadrature sum of the Poisson uncertainty related to the measured flux and the background uncertainty from the RMS image derived above.
The Kron flux is consistent with the flux estimated from the growth curve,
$f_{G_{\rm W}, \rm cog} = 6.0 \pm 0.6 \times 10^{-20}$~erg~s$^{-1}$~cm$^{-2}$~$\AA^{-1}$, indicated in yellow in Fig.~\ref{fig:cog_GW}.

We repeat the analysis for $G_{\rm E}$, using both the Kron flux and the growth curve method. The purple line in Fig.~\ref{fig:cog_GE} marks the total flux evaluated from the elliptical Kron aperture ($f_{G_{\rm E}, \rm Kron} = 7.1 \pm 2.5 \times 10^{-21}$~erg~s$^{-1}$~cm$^{-2}$~$\AA^{-1}$).
Source $G_{\rm E}$ is much fainter and compact, and the resulting growth curve is noisier.
For estimating the total flux, we, therefore, apply an aperture correction relying on the values published in the WFC3/IR handbook.
The flux values in the apertures encircling 70.8\%, 83.6\%, 86.3\%, and 90\% of the total energy are shown as colored squares in Fig.~\ref{fig:cog_GE}. At radii $\gtrsim 0\farcs4$ the noise becomes considerable. 
We hence rely on the aperture at $0\farcs25$ (blue square), obtaining a total aperture-corrected flux $f_{G_{\rm E}} = 8.3 \pm 2.4 \times 10^{-21}$~erg~s$^{-1}$~cm$^{-2}$~$\AA^{-1}$.

The Kron and growth curve fluxes are consistent with each other at 1$\sigma$ level, as the substantial uncertainties easily accommodate the $\approx10\%$ difference.
In the following, we adopt as best estimate the value of the flux obtained from the growth curve.
The apparent magnitude of $G_{\rm E}$ thus becomes $m(AB)_{G_{\rm E}} = 26.9 \pm 0.2$~mag.
Information relative to the two sources is summarized in Table~\ref{tab:twogal}.

\begin{table*}
    \caption{Properties of galaxies $G_{\rm W}$ and $G_{\rm E}$}
    \centering
    \begin{tabular}{c c c c c c}
        \hline\hline
       Source  & $d_{\rm quasar}$ & F160W Flux & F160W mag & SFR & M$_*$ \\
        & (kpc) & (erg s$^{-1}$ cm$^{-2}$ $\rm \AA^{-1}$) & (AB mag) &( M$_\odot$ yr$^{-1}$) & (M$_\odot$) \\
       \hline
       $G_{\rm W}$ & $19.5$ & $60 \pm 4 \times 10^{-21}$ & $24.71 \pm 0.08$ & $5.0 \pm 0.4$ & $\lesssim 9.9 \times 10^9$\\
       $G_{\rm E}$  & $40$ & $8.3 \pm 2.4 \times 10^{-21}$ & $26.9 \pm 0.3$ & $0.70 \pm 0.20$ & $\lesssim 1.4 \times 10^9$ \\
       \hline
    \end{tabular}
    \tablefoot{From left to right: name of the source; the projected distance with respect to the quasar ($d_{\rm quasar}$); the observed F160W flux; the apparent AB magnitude in the F160W filter; the star formation rate evaluated with \textsc{Starburst99} assuming a constant star formation history; the inferred stellar mass.}
    \label{tab:twogal}
\end{table*}

\subsection{Star formation rate and stellar mass evaluation}

\begin{figure*}
    \centering
    \begin{subfigure}{0.45\textwidth}
        \includegraphics[width=\textwidth]{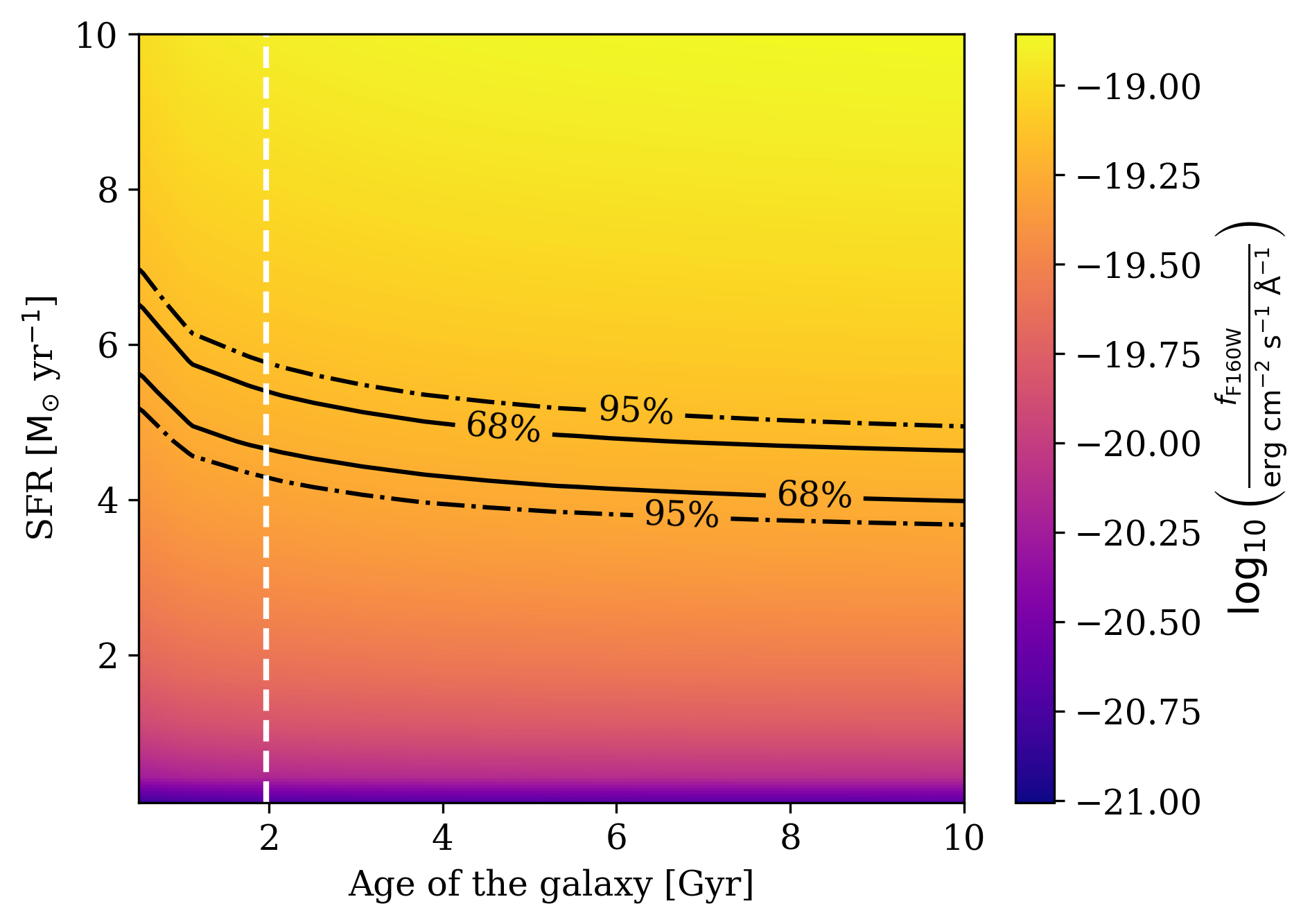}
        \caption{G$_{\rm W}$}
        \label{fig:GW_simflux}
    \end{subfigure}
    \begin{subfigure}{0.45\textwidth}
        \includegraphics[width=\textwidth]{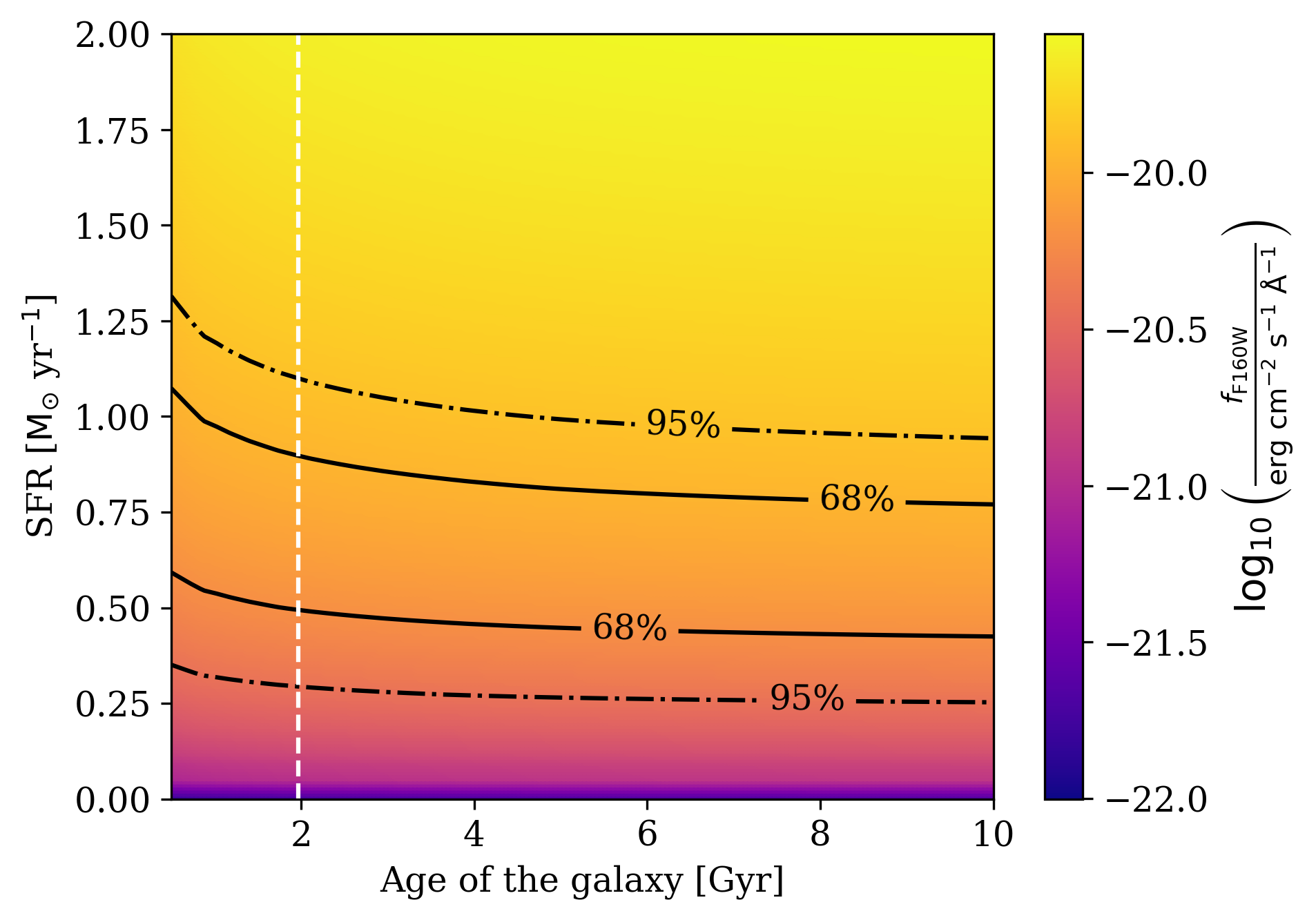}
        \caption{G$_{\rm E}$}
        \label{fig:GE_simflux}
    \end{subfigure}
    \caption{(a) F160W simulated flux of a $z=3.25$ galaxy as a function of SFR and age, assuming a constant SFR evolution. The black solid and dash-dotted lines are the 1$\sigma$ and 2$\sigma$ confidence intervals on the Kron flux of $G_{\rm W}$. The white line marks the age of the Universe at $z = 3.25$, i.e. $t = 1.97$ Gyr, in correspondence of which the average star formation rate of galaxy $G_{\rm W}$ is inferred to be SFR$_{G_{\rm W}} = 5.0 \pm 0.4$ M$_{\odot}$~yr$^{-1}$. (b) Same as left panel but for galaxy $G_{\rm E}$.}
    \label{fig:simflux}
\end{figure*}

\begin{figure}
    \centering\includegraphics[width=0.9\columnwidth]{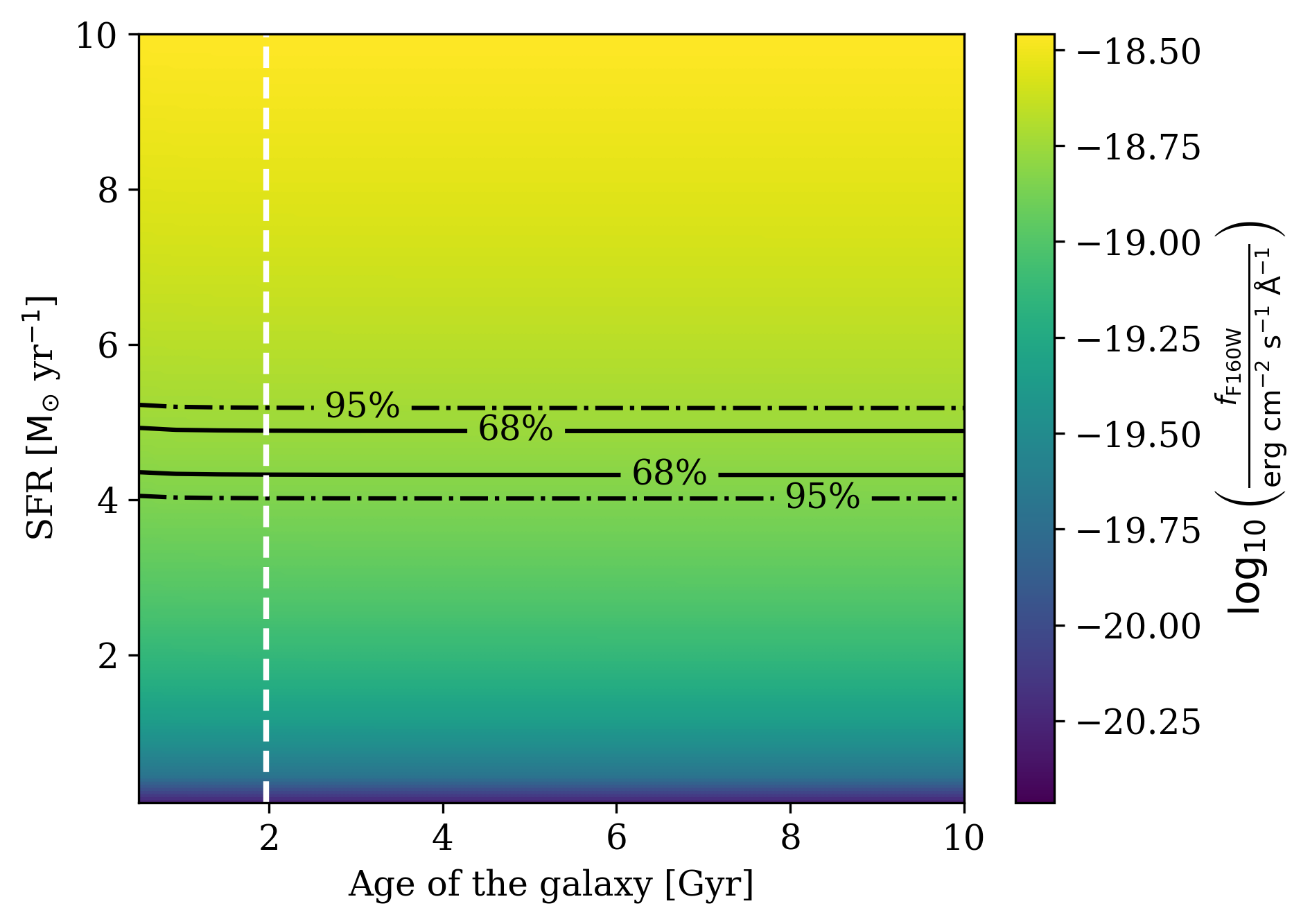}
    \caption{Simulated $r-$band flux as a function of the galaxy age and SFR, assuming a constant star formation history. The black solid and dashed lines are the 1$\sigma$ and 2$\sigma$ confidence intervals of the $r$ flux estimation measured with MUSE data by \citet{Fumagalli2017}. The white dashed line marks the age of the Universe at $z = 3.25$, i.e. $t = 1.97$ Gyr. The $r$-band flux is more sensible to the instantaneous SFR, and under the assumption of constant star formation history used in this case, it does not depend on the age of the galaxy.}\label{fig:simflux_r}
\end{figure}

\begin{figure}
    \includegraphics[width=0.8\columnwidth]{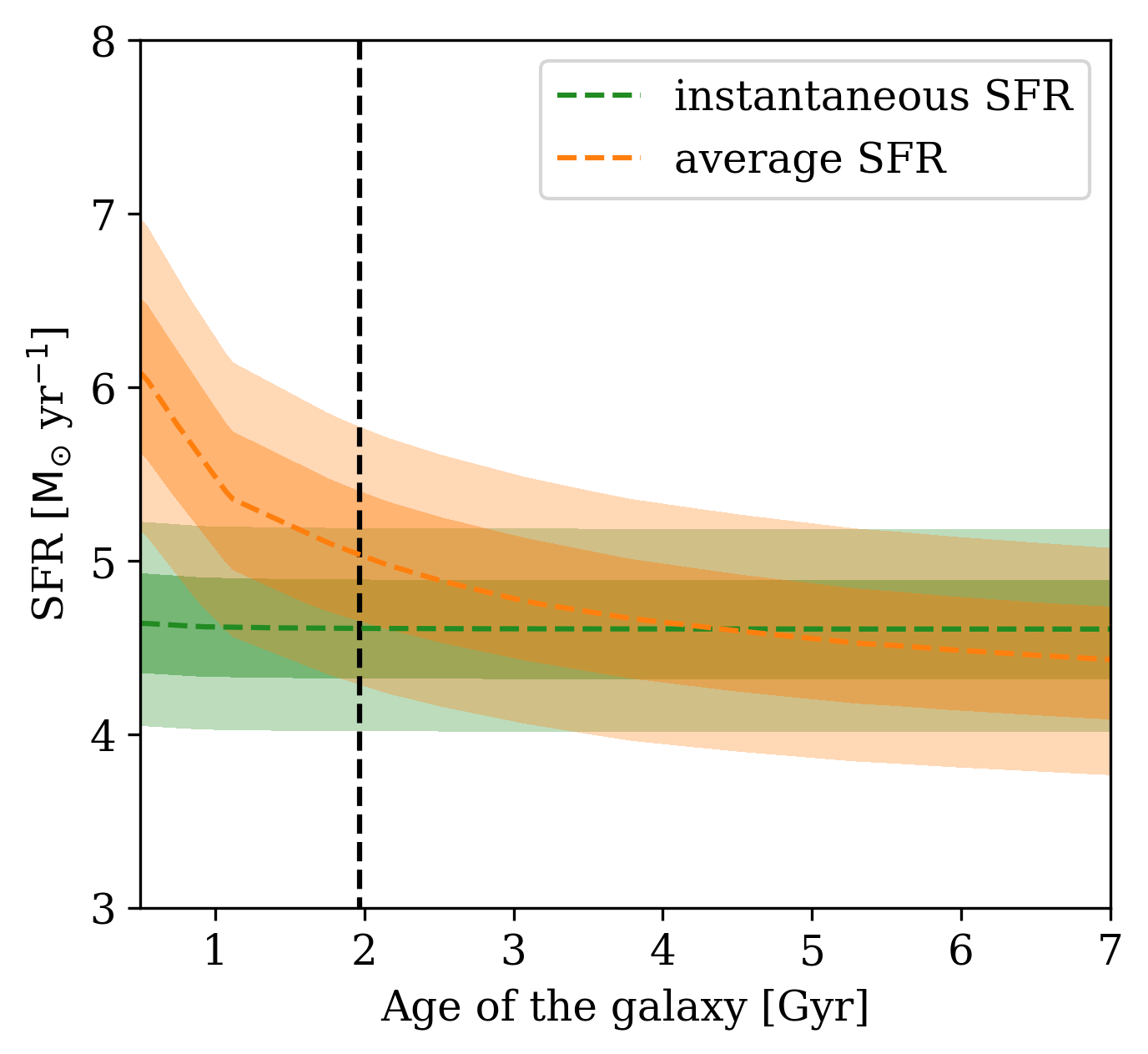}
    \caption{
    Comparison between the inferred instantaneous SFR (green) inferred from the $r$-band flux and the value averaged over longer timescales (orange) from the F160W photometry of galaxy $G_{\rm W}$.
    The dark and light-shaded regions mark the 68\% and 95\% confidence intervals. The similarity between the two values corroborates the hypothesis of constant SFR.}
    \label{fig:GW_comp_r}
\end{figure}

To evaluate the stellar mass of the two sources we simulate the spectral evolution using the \textsc{Starburst99} software \citep{Leitherer1999}, assuming Geneva stellar tracks without rotation and a metallicity comparable to that of the DLA, $Z=0.002$ ($Z \simeq 0.1 Z_\odot$). 
We assumed a Kroupa \citep{Kroupa2001} initial mass function, represented by a double power-law with index $\alpha_1 = 1.3$ between $0.1-0.5$ M$_\odot$ and $\alpha_2 = 2.3$ between $0.5-100$ M$_\odot$. Dust is not considered by \textsc{Starburst99} models, however, even if dust is not expected to significantly affect LAEs, we discuss the consequences of the presence of dust at the end of this section.
Due to the limited number of photometric data points available, we are unable to accurately constrain all the necessary properties, such as dust content, age and star formation history, required to estimate the stellar mass of the galaxies. As a result, we must proceed with our analysis by introducing assumptions, some of which are testable, with the aim of establishing an upper limit for the stellar mass of the two sources.

To explore the properties of $G_{\rm W}$ we consider the evolution of a galaxy with constant SFR ranging from 0.1 to 10~M$_{\odot}$~yr$^{-1}$, evaluating the flux that would produce in the F160W filter if it were observed at the DLA redshift $z_{\rm dla}=3.2552$. The results are shown in Fig.~\ref{fig:GW_simflux}, in which the color illustrates the value of the synthetic IR flux as a function of SFR and age. 
The black solid and dash-dotted lines enclose the 68\% and 95\% confidence intervals around the measured value of $f_{G_{\rm W}, \rm Kron}$. 
We observe that the SFR implied by the F160W magnitude strongly depends on age for ages $<2$ Gyr, but it is relatively constant thereafter.
This trend arises from the fact that the F160W band has a pivot wavelength at $\lambda_{\textup{F160W}} = 15369$ $\AA$, corresponding to a rest frame emission of $\lambda_{\textup{rf}} \simeq 3612$ $\AA$, i.e., at the boundary between the near UV (NUV) and optical bands.
The NUV part of the spectrum is more sensitive to the SFR of the galaxy, but the probed wavelengths are red enough to be also affected by the growth of the stellar mass. Due to the additional contribution from older stellar populations at a later time, the same flux can be obtained with a lower SFR than at earlier times. 
To find an upper limit for the mass of the galaxy, we formulate the hypothesis that the age of the galaxy is approximately equal to the age of the Universe at the observed redshift, i.e., $t \simeq 1.97$ Gyr, marked by the white vertical line in Fig.~\ref{fig:GW_simflux}, deriving an average SFR$_{G_{\rm W}} = 5.0 \pm 0.4$~M$_{\odot}$~yr$^{-1}$.

The assumption of constant star formation history for galaxy $G_{\rm W}$ can be tested using the observed $r$-band flux of $f_{r} = 16 \pm 1 \times 10^{-20}$~erg~cm$^{-2}$~s$^{-1}$~$\AA^{-1}$
computed by \cite{Fumagalli2017} through the convolution of MUSE spectroscopic data with the $r$ filter response. 
The observed $r$-band flux, with a pivot wavelength $\lambda_{r} \approx 6176$~$\AA$, is more sensitive to the instantaneous SFR of the galaxy as it is probing the rest frame far UV (FUV) flux with $\lambda_{r, \textup{rf}} \approx 1451$~$\AA$.
As before, we evaluate the synthetic $r$-band flux that galaxies with different values of SFR should emit and compare them with what observed for galaxy $G_{\rm W}$. 
The results showed in Fig.~\ref{fig:simflux_r} demonstrate how the $r$-band flux can be considered independent from the age of the galaxy under the assumption of constant star formation history and the instantaneous SFR of galaxy $G_{\rm W}$ is measured to be SFR$_{G_{\rm W}} = 4.6 \pm 0.3$ M$_\odot$ yr$^{-1}$.
In Fig.~\ref{fig:GW_comp_r}, we compare the more instantaneous SFR value inferred from the $r$-band flux (green line) with the 
indicator on longer time scales from the F160W photometry (orange line). 
The shaded regions mark the 1$\sigma$ and 2$\sigma$ confidence intervals on the measured fluxes. 
The fact that at $t \simeq 1.97$ Gyr, the assumed age of $G_{\rm W}$, the instantaneous SFR is comparable with the averaged SFR obtained from F160W observations rules out 
significant excursions in the recent star formation history of the galaxy. 
The slightly lower value of the inferred instantaneous SFR could be explained by a past enhancement in the star formation activity of $G_{\rm W}$. Still, the difference is within current uncertainties.

The strawman assumption of $t \simeq 1.97$ Gyr can be compared to what is found in the stellar populations of LAEs. For example, recent studies suggest that LAEs have typical ages of $\lesssim 500$ Myr \citep[e.g.,][]{Matthee2021, Endsley2024}. 
If that were the case, we would derive an average SFR for $G_{\rm W}$ of $\gtrsim 6.1$ M$_\odot$ yr$^{-1}$, a factor 1.3 higher than the instantaneous star formation rate, suggesting that the galaxy may have undergone a more significant starburst phase.
Assuming a maximal age also provides an estimate for the galaxy maximum stellar mass using the relation:
\begin{equation}
    {\rm M_*} = \int_0^{t = 1.97 {\rm Gyr}} {\rm SFR_{G_{\rm W}}} \ dt 
    \end{equation} 
This leads to a maximum stellar mass of $M_{G_{\rm W}} < 9.9 \pm 0.7 \times 10^9$~M$_{\odot}$, consistent with the mass $\approx 5 \times10^9$ M$_\odot$ estimated through scaling relations calibrated on known DLAs \citep{Moller2013}.
If the age of the galaxy was around $500$ Myr then the same argument would lead to $M_{G_{\rm W}} \approx 3.1 \times 10^{9}$ M$_\odot$.

We repeat the analysis for source $G_{\rm E}$
and show the results in Fig.~\ref{fig:GE_simflux}. 
The same behavior described in Fig.~\ref{fig:GW_simflux} for $G_{\rm W}$ can also be observed in this case, where the increment in time of the stellar mass results in an increment of the F160W mock flux at constant SFR.
Assuming that also the age of $G_{\rm E}$ is comparable with the age of the Universe at the DLA redshift, we estimated the SFR of the source to be SFR$_{G_{\rm E}} = 0.70 \pm 0.20 $~M$_\odot$~yr$^{-1}$, leading to a maximum stellar mass of M$_{G_{\rm E}} < 1.4 \pm 0.4 \times 10^9$~M$_\odot$, about one order of magnitude lower than M$_{G_{\rm W}}$.

We do not have any other observation to test the hypothesis of a constant SFR for galaxy $G_{\rm E}$.
However, when using \textsc{Starburst99} to generate mock F160W fluxes in the case of starburst, we find that to reproduce the observed photometry, the source would need to have a stellar mass $10^{9.4} < M_{G_{\rm E}}< 10^{9.7} \ \rm M_\odot$ when we require the galaxy to have an age between $0.5 - 1$ Gyr.
A single starburst with such a high mass is very unlikely, and therefore, we consider it more probable that this galaxy has a constant SFR. 
However, we do not exclude the possibility that the star formation history of $G_{\rm E}$ could be variable, with alternate phases of intense and moderate activity as observed for stochastic star formation in low-mass systems \citep{Fumagalli2011,Guo2016}.

The SFR values that we have considered until now are the ``observed'' SFR, i.e., they are not corrected for dust reddening. With the limited photometric information we currently have, we cannot derive the dust properties independently.
Therefore, to estimate the impact of dust, we can rely on the values observed in other LAE samples.  \citet{Matthee2021} find that LAEs at $z \sim 2$ have similar SFR to $G_{\rm W}$ ($\approx 6 \pm 1$ M$_\odot$ yr$^{-1}$) and an extinction of $A_V \approx 0.7$.
This extinction would lead to a factor $1.9$ higher intrinsic flux. This translates into an intrinsic SFR$_{G_{\rm W}} \approx 9.5$ M$_{\odot}$ yr$^{-1}$ and SFR$_{G_{\rm E}} \approx 1.33$ M$_{\odot}$ yr$^{-1}$.
Since the dust correction is not well characterized, we will rely on the minimum observed SFR value in the following. As we will show, even without considering dust correction, the ionizing radiation emitted by the two sources is sufficient to power the observed Ly$\alpha$ luminosity, and dust correction will only strengthen this argument.


\section{Discussion}
\label{sec:discussion}

\subsection{What powers the  Ly$\alpha$ emission?}
\label{sec:lyaorigin}

Constraining the physical processes at the origin of Ly$\alpha$ emission in galaxies is challenging since many plausible powering mechanisms have been suggested \citep[e.g.,][]{Ouchi2020}. The observed radiation is likely generated by a combination of them, as also shown by simulations \citep[e.g.,][]{Mitchell2021}.
One possibility is that Ly$\alpha$ photons are produced {\it in-situ} in the CGM from fluorescence after photoionization. Alternatively, radiation can be powered by cooling during the accretion of gas onto dark matter halos.
Ly$\alpha$ photons can also be generated in \ion{H}{ii} regions of galaxies and then diffused in the CGM through scattering on the surface of neutral gas clouds. Many studies propose that this scattering process constitutes the primary mechanism, although they cannot definitively rule out the contribution of other processes \citep[][]{Wisotzki2016, Leclercq2017}. 
The hypotheses about Ly$\alpha$ photons generated by recombination following photoionization in the CGM or in the ISM are highly plausible from an energetic point of view, even though the expected emission from theoretical models usually suffers from high uncertainties due to the unknown geometry of the neutral gas clouds.
However, this is a central factor in shaping the observed radiation and understanding the amount of ionizing or Ly$\alpha$ photons able to escape and generate the nebula.
Moreover, some Ly$\alpha$ photons might be produced by UV-faint undetected satellites surrounding the central galaxy, believed to play a central role in explaining extended ($\gtrsim 50$ kpc) asymmetric Ly$\alpha$ emission \citep[e.g.,][]{HerreroA2023}.

With HST NIR observations, we have discovered the presence of a new UV emitting source, $G_{\rm E}$, inside the Ly$\alpha$ emitting clump $C_{\rm E}$. No other sources are detected within the nebula at the current depth, except the previously known galaxy $G_{\rm W}$ that partially overlaps with the clump $C_{\rm W}$.
With these data, we only wish to answer whether the UV photons produced by the two detected galaxies are enough to power the Ly$\alpha$ nebula via {\it in-situ} photoionization?
\citet{Fumagalli2017} found that galaxy $G_{\rm W}$ alone was only marginally able to power the entire nebula, making it more plausible that other sources contribute to the ionization. 
We now reassess the hypothesis of photoionization as a powering mechanism from an energetic point of view, considering both sources.

Under case-B recombination, the Ly$\alpha$ luminosity and the rate of hydrogen ionizing photons emitted by a galaxy are related by 
\begin{equation}
    L_{\mathrm{Ly\alpha}} = 0.68 \ f_{\rm esc, LyC}\  Q_{\mathrm{HI}} \ E_{\mathrm{Ly\alpha}}\:,
\end{equation}
where $E_{\mathrm{Ly\alpha}}$ is the energy of Ly$\alpha$ photons, the factor $0.68$ denotes the probability under case-B that a recombination event will result in the emission of a Ly$\alpha$ photon \citep{Dijkstra2014},
and $Q_{\rm{HI}}$ represents the photoionization rate of the galaxy given its star formation.
$f_{\rm esc, LyC}$ is a multiplicative factor that accounts for the fraction of ionizing photons that can leave the interstellar medium and ionize the surroundings. 
As introduced above, there is a plethora of possible mechanisms able to power Ly$\alpha$ emission, and we do not exclude that some of them might play an important role even in our case. For this reason, the values of $f_{\rm esc, LyC}$ that we present under the hypothesis that Ly$\alpha$ is only due to photoionization in the CGM should be interpreted as maximum escape fractions, i.e., the contribution needed to power the nebula {\it in-situ}.
In the following, we use the results of the \textsc{Starburst99} models described above to estimate the ionizing photon rate. 

Based on the new estimates of SFR in Table~\ref{tab:twogal} for the two sources in our field, we obtain $Q_{\mathrm {HI}, G_{\rm W}} = (1.44 \pm 0.12) \times 10^{54}$~s$^{-1}$ and $Q_{\textup{HI}, G_{\rm E}} = (2.0 \pm 0.6) \times 10^{53}$~s$^{-1}$.
Cumulatively, the ionizing photon rate of $\approx 1.6\times 10^{54}$~s$^{-1}$ is sufficient to power the total observed Ly$\alpha$ luminosity of $L_{\rm Ly\alpha, neb} = (2.7 \pm 0.1) \times 10^{42}$~erg~s$^{-1}$ if only $\approx 15\%$
of the ionizing photons escape from the galaxies ($f_{\rm esc, LyC} = 0.15$). Hence, the detection of an additional galaxy $G_{\rm E}$ and a revised estimate of the SFR for galaxy $G_{\rm W}$ compared to \cite{Fumagalli2017} is enough to power the Ly$\alpha$ nebula.

Considering the two galaxies individually, the maximum Ly$\alpha$ luminosity that we can expect to observe under our hypothesis is the one obtained imposing the maximum escape fraction $f_{\rm esc, LyC} = 1$, i.e., $L_{\rm Ly\alpha, G_{\rm W}}(f_{\rm esc, LyC}=1) = (1.60 \pm 0.13) \times 10^{43}$~erg~s$^{-1}$ and $L_{\rm Ly\alpha, G_{\rm E}}(f_{\rm esc, LyC}=1) = (2.2 \pm 0.6) \times 10^{42}$~erg~s$^{-1}$. If we attribute the photons escaping each source to the powering of each clump separately, we can estimate an escape fraction needed for each galaxy.
To do so, we compare maximum luminosities above with the values observed in \citet{Fumagalli2017} for clump W, $L_{\rm Ly\alpha,C_{\rm W}}=(9.5 \pm 1.2) \times 10^{41}$~erg~s$^{-1}$, and clump E, $L_{\rm Ly\alpha,C_{\rm E}} = (7.6 \pm 0.8) \times 10^{41}$~erg~s$^{-1}$.
Considering galaxy $G_{\rm E}$ fully embedded within the clump, the inferred escape fraction is $ f_{\rm esc, LyC} \approx 0.34$. For galaxy $G_{\rm W}$, instead, as argued in \cite{Fumagalli2017}, the spatial offset between the source and the clump can be interpreted with a geometry in which galaxy $G_{\rm W}$ is not fully embedded within the emitting structure, but is shining towards the clump that receives $\approx 1/4$ of the total photon flux.
In this case, the escape fraction becomes $f_{\rm esc, LyC} \approx 0.24$.

These escape fraction values are higher than the average $f_{\rm esc, LyC}$ for LAEs at similar redshift, which is observed to be $\lesssim 10 \%$ \citep{Marchi2017, Naidu2018}, but still represent feasible values \citep[e.g.,][]{Naidu2022}.
For this reason, we argue that {\it in-situ} photoionization is able to power the Ly$\alpha$ nebula, although it is reasonable that Ly$\alpha$ photons generated in the ISM of galaxies and scattered outward contribute as well. Furthermore, as discussed in Section~\ref{sec:analysis}, if the galaxies were younger or if the absorption due to dust is not negligible, the SFR values would be higher and lower escape fractions would be required to explain the observed Ly$\alpha$ emission. Moreover, if $G_{\rm E}$ with its estimated SFR and its position at the center of the clump $C_{\rm E}$ is arguably powering the left clump of the nebula, because of the offset between galaxy $G_{\rm W}$ and clump 
$C_{\rm W}$, we do not exclude the possibility that a further low-mass galaxy below the detection limit is part of this group and adds additional photons to power the Ly$\alpha$ line by in situ star formation. If such a source were present in the region of the nebula, it would have a maximum flux of $\lesssim 6.16 \times 10^{-21}$ erg cm$^{-2}$ s$^{-1}$ \AA$^{-1}$, as inferred from the 3$\sigma$ detection level in NIR/HST observations. The SFR of a further undetected source would be $\lesssim 0.5$ M$_{\odot}$ yr$^{-1}$, and it would provide a rate of ionising photons of $\lesssim 1.5 \times 10^{53}$ s${^{-1}}$. Therefore, a single source near the detection limit, or multiple fainter sources, can contribute to the overall Ly$\alpha$ luminosity. However, the presence of the two galaxies $G_{\rm W}$ and $G_{\rm E}$ alone provides enough ionizing photons to generate the observed Ly$\alpha$ luminosity.

\subsection{A mid-size nebula in an \ion{H}{I} rich galaxy group}

The discovery of a new galaxy, $G_{\rm E}$, through HST imaging, strengthens the argument presented in previous analysis that the DLA arises from an extended structure associated with a galaxy group. Under the assumption that galaxy $G_{\rm W}$ is the group central galaxy, known scaling relations would imply a maximum halo mass for the group of $\approx 10^{12}$~M$_\odot$ \citep{Moster2010}. Extrapolating scaling relations from simulations \citep{Evrard2008, Munari2013}, groups in this mass range would have a one-dimension velocity dispersion of $\approx 150 - 165$~km~s$^{-1}$. 
From the study of the profiles of the DLA metal lines, a velocity dispersion of the order of $\approx 150$~km~s$^{-1}$ is found \citep{Fumagalli2017}, consistent with the inferred group mass, when assuming that the metal-rich gas clouds are moving in the same potential of the galaxies.

Given the configuration of the two galaxies, the brighter Ly$\alpha$ emission can be interpreted as halo gas illuminated by ionizing photons, as argued above. The fainter, more extended emission can instead be associated with an intragroup medium. As the two galaxies are separated by only $\approx 30$~kpc in projection and hence interacting, it is quite likely that tidal material is present inside the group. Similar features have been observed at lower redshifts inside groups \citep{Chen2019,Leclerq2022}. At higher redshift, only a handful of systems in which Ly$\alpha$ emission associated with a DLA suggests the presence of a group environment \citep{Moller2002, Weatherley2005, Fynbo2023} have been detected. The presence of substantial amounts of neutral gas at a projected impact parameter of $\approx 20$~kpc in the south-west direction further confirms that the emission is tracing only a portion of the group environment, which is likely to extend over an area of more than $30\times 50$~kpc. 
Hence, this mid-size nebulae is the signpost of moderate-mass groups in the middle of the range bracketed by the Ly$\alpha$ halos of star-forming galaxies and the extended quasar nebulae. 
However, the lack of a significant number of such extended structures reported in the literature, particularly in light of blind and systematic searches near absorbers \citep[e.g.][]{Lofthouse2023,Galbiati2023}, implies that these are rarely seen in emission at $z\gtrsim 3$.

\subsection{Model predictions}

\citet{Krogager2017} developed a model based on the earlier work by \citet{Fynbo2008} to predict the UV luminosity of possible galaxies associated to DLAs. Given our DLA metallicity and impact parameter of the galaxy $G_{\rm W}$, this model predicts an absolute UV magnitude at 1700\AA\ of -18.75. When we estimate the observed UV magnitude from the $r$-band flux we obtain -20.09. We observe a magnitude of a factor $\sim 1$ lower than the predicted one, and in figure 11 from \citet{Krogager2017}, our source would lie in the region where other DLA-associated galaxies at $z\gtrsim 2$ have been detected \citep{Christensen2014}. 

Using the EAGLE (Evolution and Assembly of GaLaxies and their Environments) cosmological simulation \citep{Schaye2015, Crain2015}, we investigate properties of galaxies similar to the sources analyzed in this work.  
When considering the observed SFR, our sources are found to form stars at a lower rate compared to galaxies with similar masses on the EAGLE main sequence, corroborating the hypothesis that the estimated stellar mass represents a higher limit and the true value is probably lower. 
We found that the expected maximum halo mass for a galaxy with a stellar mass similar to $G_{\rm W}$ should be $\sim 10^{12}$ M$_{\odot}$, as also inferred by known scaling relations. Similar galaxies usually host a central massive black hole, so we do not exclude the possibility that some of the ionizing photons could be generated through accretion onto the black hole. However, galaxies in the simulation are usually found with a higher metallicity ($Z \gtrsim Z_{\odot}$) than the assumed one, but given the associated scatter, this difference is unlikely to play a key role in our interpretation.
Our observations also offer a test for the analytic model by \citet{Theuns2021}, who predicted the properties of DLAs in the cold dark matter model. We find that given the impact parameter of $G_{\rm W}$ and the halo mass, the predicted column density of the DLA is $N_{\rm HI} \approx 10^{21}$ cm$^{-2}$, at less than 2$\sigma$ from the observed value reported in \citet{Fumagalli2017}. 
However, accounting for a slightly lower halo mass would provide a column density more consistent with the observed one, again underlining that the halo mass we estimated represents a higher limit.

\section{Summary and Conclusions}
\label{sec:conclusions}

We present WFC3/IR observations of a double-clumped emitting Ly$\alpha$ nebula ($L_{\rm Ly\alpha} = 27 \pm 1 \times 10^{41}$~erg~s$^{-1}$) that has been detected in correspondence of a DLA with column density 
$\log (N_{\textsc{hi}}/\rm cm^{-2}) = 20.85 \pm 0.10$ and metallicity 
$\log Z/Z_\odot = -1.1 \pm 0.1$ at redshift $z_{\rm dla} = 3.2552 \pm 0.0001$. Previous observations highlighted how gas in absorption is connected to the Ly$\alpha$ emission due to the similar line-of-sight velocity difference between the two emitting clumps and the two components in the DLA metal lines.

With new F160W observations, we unveiled the presence of a faint galaxy $G_{\rm E}$, whose center is at a projected distance $< 2$ kpc from the Ly$\alpha$ emitting clump $C_{\rm E}$, and hence is considered embedded in the nebula. In addition, we measured the rest-frame continuum at 3600~\AA\ of the previously-known galaxy $G_{\rm W}$ \citep[called $G$ in][]{Fumagalli2017}, which is offset by $\approx 7.7$ kpc from clump $C_{\rm W}$ and is consistent with being an LBG with $z \sim 3.2$.
Hence, we conclude that the two continuum sources are physically associated with the nebula. 

Combining photometry from HST and  \textsc{Starburst99} models, we inferred an SFR$_{G_{\rm W}} = 5.0 \pm 0.4 \ \rm M_\odot$~yr$^{-1}$ and SFR$_{G_{\rm E}} =  0.70 \pm 0.20 \ \rm M_\odot$~yr$^{-1}$ under the assumption of constant star formation history and considering the age of the sources comparable to the age of the Universe at the DLA redshift ($t = 1.97$ Gyr). 
When applying typical values of dust extinction observed for DLA-associated galaxies at a similar redshift the intrinsic SFR can rise up to SFR$_{G_{\rm W}} \approx 9.5$ M$_{\odot}$ yr$^{-1}$ and SFR$_{G_{\rm E}} \approx 1.33$ M$_{\odot}$ yr$^{-1}$. 
We also obtain an estimate of the maximum stellar masses integrating the observed SFR values, equal to M$_{G_{\rm W}} = 9.9 \pm 0.7 \times 10^9$ M$_\odot$ and M$_{G_{\rm E}} = 1.4 \pm 0.4 \times 10^9$ M$_\odot$, respectively.
For galaxy $G_{\rm W}$, the assumption of a constant star formation history is corroborated by a consistent value for the instantaneous SFR, obtained from the rest-frame FUV from \citet{Fumagalli2017}.
For galaxy $G_{\rm E}$, undetected in the $r-$band, the evolution of the galaxy under starburst condition would require a mass of $10^{9.4} < M_*/M_\odot < 10^{9.7}$ for a burst age of $0.5 - 1$~Gyr, which we consider very unlikely for a single burst.
However, the age of the galaxies still remains unconstrained, and observational studies suggest that LAEs at these redshifts are usually younger. 
In this case, the galaxies would display a higher SFR but a lower stellar mass (if $t=500$ Myr we would obtain SFR $\gtrsim 6.1$ M$_\odot$ yr$^{-1}$, M$_{G_{\rm W}} \approx 3.1 \times 10^9$ M$_\odot$).

With the revised values of star formation rates and the detection of a new source, we revisit whether photoionization in the CGM due to the two galaxies alone could power the nebula. This option is viable if $\approx 15\%$ of the ionizing photons leave the galaxies and are absorbed inside the nebula. Considered individually, both galaxies can power the respective clumps with escape fractions of $0.24$ and $0.34$ for galaxies $G_{\rm W}$ and $G_{\rm E}$, respectively. 
These values, even if similar to some observations, are above the average of $f_{\rm esc, LyC}$, and for this reason we do not exclude the possible presence of additional sources with individual observed SFR of $\lesssim 0.5$ M$_\odot$ yr$^{-1}$ that could contribute to the number of energetic photons and additional Ly$\alpha$ powering mechanisms inside the CGM and within groups, or additional contributions of Ly$\alpha$ photons scattered from the ISM.

The detection of a new source strengthens the argument that this DLA originates from a galaxy group, for which we inferred a maximum halo mass of $\approx 10^{12}$~M$_\odot$ from both the derived stellar mass and the DLA kinematics. Also, the EAGLE cosmological simulation predicts a similar halo mass for galaxy $G_{\rm W}$, but a higher metallicity ($Z \gtrsim Z_{\odot}$). The estimated upper limits on the properties of galaxy $G_{\rm W}$ seem to fit consistently in the landscape of both analytical \citep{Theuns2021} and empirical \citep{Krogager2017} models.

Given these pieces of information, we deduce that the Ly$\alpha$ emission traces the intergalactic medium and material displaced by the interacting galaxies in the inner part of the group. At the same time, the neutral hydrogen detected at the projected quasar position shows how the emission is tracking only a portion of a structure that can extend for more than $30 \times 50$~kpc. Although such groups should be generally common already at $z\approx 3$, the apparent scarcity of mid-size Ly$\alpha$ nebulae, like the one we investigated in this work, seems to suggest a gap in the distribution of sizes between individual galaxies ($\lesssim 30$~kpc) and bright quasar hosts ($\gtrsim 80-100$~kpc). Dedicated searches for mid-sized ($\approx 40-60$~kpc) nebulae will better constrain their number density and the degree of association with galaxy groups.

\begin{acknowledgements}
We thank Lise Christensen, Palle M{\o}ller, Johan Fynbo, and the anonymous referee for constructive comments that helped to improve this work. Support for Program 15283 was provided by NASA through grants from the Space Telescope Science Institute, which is operated by the Association of Universities for Research in Astronomy, Incorporated, under NASA contract NAS5-26555. This project has received funding from the European Research Council (ERC) under the European Union's Horizon 2020 research and innovation program (grant agreement No 757535), by Fondazione Cariplo (grant No 2018-2329), and is supported by the Italian Ministry for Universities and Research (MUR) program' Dipartimenti di Eccellenza 2023-2027', within the framework of the activities of the Centro Bicocca di Cosmologia Quantitativa (BiCoQ).
\end{acknowledgements}


\end{document}